**Hybrid superconducting-magnetic memory device using competing order parameters[a]**

Burm Baek, William H. Rippard, Samuel P. Benz, Stephen E. Russek, and Paul D. Dresselhaus

National Institute of Standards and Technology, 325 Broadway, Boulder, CO 80305, USA

**Abstract**

Superconducting devices, which rely on modulating a complex superconducting order parameter in a Josephson junction, have been developed for low power logic operations, high-frequency oscillators, and exquisite magnetic field sensors. Magnetic devices, which rely on the modulation of a local vector order parameter- the local magnetic moment, have been used as memory elements, high-frequency spin-transfer oscillators, and magnetic field sensors. In a hybrid superconducting-magnetic device, these two order parameters compete, with one type of order suppressing the other. Recent interest in ultra-low-power, high-density cryogenic memories has spurred new interest in merging superconducting and magnetic behavior so as to exploit these two competing order parameters to produce novel switching elements. Here, we describe a reconfigurable two-layer magnetic spin valve integrated within a Josephson junction. Our measurements separate the suppression in the superconducting coupling due to the exchange field in the magnetic layers, which causes depairing of the supercurrent, from the suppression due to the magnetic field generated by the magnetic layers. The exchange field suppression of the superconducting order parameter is a tunable and switchable behavior that is also scalable to nanometer device dimensions. These devices are the first to demonstrate nonvolatile, size-independent switching of the Josephson coupling, in both magnitude and phase, and they may allow for the first nanoscale superconducting memory devices.

Superconducting and magnetic devices have a long but mutually exclusive history due to the fact that the order parameters are competing, magnetic order suppresses superconducting order and vice versa. Superconducting Josephson junctions, which rely on modulating the magnitude and phase of the superconducting order parameter $\psi(\vec{r}) = ne^{i\theta}$, have been developed for microwave oscillators, voltage standards, logic gates, and sensitive magnetic field detectors[1]. Magnetic devices, which rely on manipulating the local magnetization $\vec{M}(\vec{r})$, have been developed for magnetic random access memory, field sensors for recording heads, and high frequency spin-transfer oscillators[2,3]. Recent advances in the understanding of superconductor-ferromagnet (S-F) hybrid structures have revealed exciting physical phenomena, such as devices in which the Josephson ground-state phase difference between the two superconductors is shifted by $\pi$ compared to that of conventional junctions, or in which Josephson coupling is achieved via a spin-polarized triplet state[4,5]. Combining the superconducting quantum and spintronic effects into low-power bi-stable devices that are switchable between different memory states could transform high-performance computing and elevate superconducting digital technology[6] as a serious alternative to existing power-hungry computers based on semiconductor technology. Despite the demonstration, over a decade ago, of a ≈700 GHz clock-rate RSFQ (rapid single flux quantum) logic element[7], the lack of a practical and scalable cryogenic memory is one reason that superconducting digital electronics has been implemented only in niche applications[8,9]. Past cryogenic-memory efforts employed circuits that stored magnetic flux quanta in superconducting loops or combined Josephson and CMOS technologies in hybrid circuits[10,11]. Unfortunately, these approaches did not simultaneously offer high-speed, ultra-low-power, and scalability.

Storing information within a Josephson junction (JJ) by changing its state is a straightforward approach to making a cryogenic memory that is both practical and scalable. One way to do this is by inserting magnetic layers within the barrier of a JJ so that the magnetic configuration changes the superconducting critical current that separates zero and nonzero voltage states[12-14]. A number of magnetically controlled Josephson switches have been demonstrated. Clinton et al.[15] demonstrated microbridge devices that switched between different critical currents using the stray field of a ferromagnet. More recently, critical-current switching was demonstrated by incorporating a single ferromagnetic layer into a JJ barrier[13]. In both devices, the difference in the critical current of the two states, or the signal contrast, comes from the magnitude and direction of the remanent magnetic field within the microbridge or the junction barrier. However, these devices will require a significant ferromagnetic moment and, thus a high magnetic switching energy at submicrometer junction dimensions, which would make them practical only for low-density memory. Scalable JJ devices should be based on direct manipulation of the Josephson coupling by use, for example, of barriers such as a pseudo-spin-valve (PSV)[12] or a multi-layer film structure with noncollinear magnetizations of the different layers for enhanced triplet coupling. Triplet superconductivity has recently become a subject of intense study due to the finite spin current and the long Cooper-pair coherence length in a ferromagnet[5,16]. Bi-stable devices appropriate for a cryogenic memory based on this effect have not been demonstrated, and their generally complicated multilayer structure and control of their noncollinear magnetic state may make such devices less practical than those based on a PSV. In this article, we focus on JJs based on PSV barriers.

A PSV comprises two different ferromagnetic layers separated by a nonmagnetic metal. Typically, its resistance state is changed through the giant magnetoresistance (GMR) effect by changing the orientation of the magnetization of one layer with respect to the other[2]. Writing information [i.e., switching the PSV state between the parallel (P) and anti-parallel (AP) magnetizations] to a typical PSV device can be accomplished either by applying a magnetic field to switch the magnetization of the layer with lower coercivity or, in nanoscale devices, by applying a bias current to switch the magnetization through the spin-transfer torque effect[17]. Regarding the superconducting transport properties of an S-PSV-S JJ, S-F proximity theory provides a physical understanding as well as a method for quantitative analysis[4,5]. The exchange field in the ferromagnet splits the two electronic spin bands, resulting in a spatial phase modulation of the Cooper-pair condensate emanating from each superconductor. This effect leads to oscillating decay, including sign reversal, of the Josephson coupling as a function of the magnetic barrier thickness. By placing a second F layer in the barrier, the total phase shift may be increased or decreased, depending on the relative orientations of the magnetizations of the F layers (i.e., P or AP), and this may produce a corresponding change in the critical current of the junction[18,19].

Experimental studies of JJs with double magnetic barriers of collinear magnetizations were carried out by Bell *et al.*[12] and later by Robinson *et al.*[20] with S-PSV-S and S-F-N-F-S (N: nonmagnetic and nonsuperconducting metal) JJs, respectively. In both studies, enhanced maximum supercurrents were observed for the AP states compared with the P states. Both works concluded that their results were due to the exchange-field effect on the pair phase, based on an argument that the phase shift in the AP state was smaller than in the P state and produced a larger, less modulated critical current. However, the interpretation of these results is complicated by competing effects. In such structures, a remanent field from the magnetic barrier induces a non-uniform supercurrent distribution within the junction and results in a maximum total supercurrent $I_m$ that is reduced from the critical current $I_c \equiv J_c A$, given by critical current density $J_c$ and the junction area $A$[21,22]. Regarding the two experiments quoted above, the higher $I_m$ in the AP state may be attributed to the lower average remanent field, as compared to that of the P state. For a rectangular junction with uniform distributions of magnetic field and $J_c$ distributions in the barrier, $I_m$ decays with increasing magnetic flux $\Phi$ in an oscillating fashion where $I_m = I_c |\sin(\pi\Phi/\Phi_0)|/(\pi\Phi/\Phi_0)$ (known as a Fraunhofer pattern[21,22]). Here, $\Phi_0$ is the magnetic flux quantum. Thus, in order to fully characterize the state of such a junction, we must extract the $I_c$, which is the maximum supercurrent at zero net flux in the barrier (not at zero applied field), determined from the maximum value of $I_m(H)$, where $H$ is the applied magnetic field.

In this work, we performed detailed $I_m(H)$ and other measurements to clearly discriminate between remanent-field effects and the more direct exchange-field effect on Josephson coupling. Distinguishing these two effects is key to understanding the physics governing these devices and evaluating their scaling potential. We found that the material choices were crucial in obtaining unambiguous results. The PSV must include materials with different coercivities, so that the device can be placed into both the P and AP states. Weaker ferromagnets make the oscillation and decay length of the Cooper pair longer and enable the use of thicker films, which is advantageous for the reproducibility of the devices. However, the free-layer coercivity must be high enough to show a large portion, including the peak, of the main lobe of the Fraunhofer pattern associated with each magnetic state of the PSV.

We used Ni as the higher-coercivity layer, since its saturation magnetization is relatively low, and at 10 K its measured coercivity is ≈ 40 mT, which is adequate for a PSV hard layer. For the free layer, we chose $Ni_{0.7}Fe_{0.17}Nb_{0.13}$, which had a coercivity of ≈ 2 mT at 10 K and a reduced magnetic moment with Nb doping[23] (Fig. 1 inset). The S-PSV-S multilayer films were sputter-deposited on oxidized silicon wafers in a chamber without breaking vacuum. Each device had the following film deposition sequence and thicknesses: Nb(100 nm)/Cu(3 nm)/$Ni_{0.7}Fe_{0.17}Nb_{0.13}$($d_{NiFeNb}$)/Cu(5 nm)/Ni($d_{Ni}$)/Cu(3 nm)/Nb(70 nm). The $Ni_{0.7}Fe_{0.17}Nb_{0.13}$ layers were grown by co-sputtering $Ni_{0.8}Fe_{0.2}$ and Nb targets. The Cu layers adjacent to each Nb layer serve as buffers or growth templates for the ferromagnetic layers and are expected to be superconducting due to the proximity effect at 4 K. The center spacer Cu layer prevents exchange coupling and reduces magnetostatic coupling between the $Ni_{0.7}Fe_{0.17}Nb_{0.13}$ and the Ni, allowing them to switch independently. Using a superconducting quantum interference device (SQUID) magnetometer, we measured the Nb superconducting temperature $T_c$ ≈ 8.9 K and observed hysteretic magnetization loops with two well-separated switching fields in the unpatterned multilayers (Fig. 1).

Four-point electrical measurements were used to characterize JJs with different dimensions and barrier materials. Wafers of test chips containing these junctions were fabricated with reliable, high-throughput processes, employing stepper lithography and reactive ion etching. The barrier etching was done by ion milling monitored with an ion mass spectrometer. The rest of the fabrication process is similar to that used to fabricate NIST Josephson voltage standards[24]. There was no noticeable deleterious change in the magnetic properties of the films due to the device processing. For junctions with designed dimensions less than 2 μm, the actual fabricated feature dimensions were significantly smaller due to process runout. For example, the smallest junctions studied were elliptical, with the short and long axes of 0.9 μm and 1.8 μm by design, but yielded effective areas $A_{eff} \approx 0.5$ μm$^2$ according to their measured resistances.

We conducted most of the electrical measurements in a liquid-helium bath at 4 K, using a dipping probe with a superconducting magnet. A magnetic field was applied parallel to the long axes of the devices. The current-voltage (*I-V*) characteristics at each magnetic field were fit to the resistively-shunted junction model[21], yielding $I_m$ and the normal resistance $R_n$ (Fig. 2c). Due to the GMR effect, the junction resistance is slightly smaller in the P state than in the AP state. The magnetoresistance ratio is roughly 0.2 % in a Ni$_{0.7}$Fe$_{0.17}$Nb$_{0.13}$(2 nm)-Ni(3 nm) PSV.

Fig. 2a shows the measured $I_m(H)$ of an S-F-S JJ (an S-PSV-S junction with an ultrathin Ni layer thickness, $d_{Ni} < 0.1$ nm) with the magnetic field swept in both directions in order to see the magnetic hysteresis. For this control sample, the Ni is so thin that it has no measureable moment and the barrier is effectively a single magnetic layer. From the positive field limit to roughly -5 mT, the magnetic state of the device stays close to a fully-magnetized, single-domain state, and the $I_m(H)$ characteristic follows a smooth Fraunhofer-like pattern. Excluding the abrupt changes, where the Ni$_{0.7}$Fe$_{0.17}$Nb$_{0.13}$ layer switches, these data can be fit by conventional theory with a horizontal shift to account for the offsets due to the net self-field of the ferromagnetic state within the barrier. We use for a perfect circular or elliptical junction[21], $I_m = I_c |2J_1(\pi\Phi/\Phi_0)|/(\pi\Phi/\Phi_0)$, where $J_1$ is a Bessel function of the first kind, to fit our data. Since $I_m(H)$ is sensitive to junction shape, $J_c$ uniformity, field uniformity, etc., this simple formula is not expected to provide a perfect fit at high fields. However, it works reasonably well for the main lobes of both PSV states. Such undistorted shapes as well as the nodes with $I_m \approx 0$ indicate no trapped flux in the junction. Each pattern is shifted in the direction opposite to the Ni$_{0.7}$Fe$_{0.17}$Nb$_{0.13}$ magnetization, as expected. The critical current $I_c$ for each magnetization state is defined by the main peak of each $I_m(H)$ for that state. The critical currents for each state are identical for this S-F-S junction, indicating no change in the Josephson coupling ($I_c$ or $J_c$), because the magnetization of the single F layer in the barrier is simply changing direction.

When adding a thicker Ni layer to form a PSV barrier, we find a different $I_c$ for each magnetic state of the S-PSV-S junction (Fig. 2b). We magnetized the Ni layer to near-saturation with $\mu_0 H \approx 200$ mT and removed the resulting trapped flux in the superconducting Nb by raising the sample temperature above 9 K. The sample temperature was then lowered to 4 K and $I_m$ was measured over a ±10 mT field range. The abrupt transitions of $I_m(H)$ indicate that the reversal of the $Ni_{0.7}Fe_{0.17}Nb_{0.13}$ layer begins at -4 mT in one direction and at +3 mT in the other direction. The two distinctly different peak values, $I_c^P = 0.11$ mA and $I_c^{AP} = 0.07$ mA for P and AP states, respectively, definitively demonstrate that the Josephson coupling can be controllably modulated by the exchange field in a PSV and that our measurements have successfully differentiated the Josephson coupling from the remanent-field effect. The remanent-field effect is significant for junction areas down to $\approx 1$ $\mu m^2$ in our devices, despite the use of weak and thin ferromagnets. However, any nanoscale device designed to exploit the remanent-field effect will suffer from a small $I_m$ modulation, because a reduced total magnetic moment and an increased demagnetizing field in the barrier result in a smaller magnetic flux trapped in the junction. The change in $I_c$, $\Delta I_c \equiv I_c^P - I_c^{AP}$, by the exchange field persists in our smallest junctions as well as in the largest ones (Fig. 3a). These results, for the first time, demonstrate the possibility of switchable nanoscale superconducting devices that may enable high-density integration for practical cryogenic memory.

Analysis of junctions with a range of $d_{Ni}$ and $A_{eff}$ provides further insight into the exchange-field effect in S-PSV-S devices. The results presented in Figs. 2b, 3a, and 3b show that $I_c^P < I_c^{AP}$, or $\Delta I_c < 0$. Since the exchange-field effect produces a phase shift of the pair wavefunction, if the thickness of one of the F layers is varied, then $\Delta I_c$ will oscillate as well, including sign changes. That is, the slope of a sinusoidal function oscillates as well as the function itself. Figs. 3e and 3f illustrate this point with the PSV barrier structure projected to a single F, that adds a phase proportional to an effective F thickness;

here, we do not take account of the phase decoherence that leads to a decay in $I_c$. The phase shift, hence the effective F thickness, is larger for the P state than for the AP state and, depending on the slope at the effective thickness of the hard layer (black dashed lines), the sign of $\Delta I_c$ can be either positive or negative. We demonstrated this behavior by fabricating and measuring JJs with different Ni thicknesses in the PSV barrier. As shown in Figs. 3c and 3d, we found two Ni thicknesses, $d_{Ni}$ = 1.5 nm and 2.0 nm, that produced very large $\Delta I_c$ with opposite signs. Such a sign change in $\Delta I_c$ is a signature of the exchange-field effect, which has not been observed to date, and shows how prominent changes in superconducting properties can result from the competition between superconducting and magnetic orders. A more complete trend has been obtained by varying $d_{Ni}$. With $0 < d_{Ni} < 4$ nm, we obtained a characteristic voltage $V_c \equiv I_c R_n$, for each state, $V_c^P$ and $V_c^{AP}$, which showed different oscillatory trends (Fig. 4). We can readily understand that $\Delta V_c$ also oscillates and changes sign. Each $V_c(d_{Ni})$ has a typical trend observed in S-F-S JJs[25,26] with different offsets in $d_{Ni}$ between the P and the AP states. Such different thickness offsets, $d_o^P$ and $d_o^{AP}$, originate from the added opposite phases with the two different magnetization orientations of the $Ni_{0.7}Fe_{0.17}Nb_{0.13}$ layer. Consequently, both $V_c^P$ and $V_c^{AP}$ can be described by an S-F-S theory for the clean limit with different $d_o^P$ and $d_o^{AP}$, respectively[27]:

$$V_c = \frac{\pi \Delta^2}{4ekT} \int_\alpha^\infty \frac{dy}{y^3} \alpha^2 \cos y. \tag{1}$$

In (1), $\Delta$ is the order parameter of the superconducting electrodes, $T$ is the temperature, and $\alpha \equiv 2E_{ex}(d_{Ni} - d_o)/\hbar v_F$ with the exchange energy $E_{ex}$ of Ni and the Fermi velocity $v_F$. We note that $\Delta$ represents a much reduced order parameter at the interfaces of Ni in this simplified model (S/Ni/S). The use of this quasiclassical theory should be appropriate for Ni thicknesses between several atomic layers and the electron mean free path (e.g., ~5 nm from Blum *et al.*[28]) and results in good fits to the measured $V_c$ data for $d_{Ni} \geq 1.5$ nm (see Fig. 4). $d_o$ includes the thin effective dead layer $d_{dead}$ in Ni; we obtained $d_{dead}$ = $(d_o^P + d_o^{AP})/2 \approx 0.8$ nm, i.e., a dead layer of 0.4 nm at each Ni/Cu interface. This is comparable to the commonly observed magnetic dead layers in F/N structures[29] and roughly consistent with our measured saturation magnetization *vs.* thickness of Ni without a $Ni_{0.7}Fe_{0.17}Nb_{0.13}$ layer. We expect a non-oscillatory decay of $V_c$ and zero $\Delta V_c$ within the dead layer if the ferromagnetism is completely suppressed[30,31]. However, the exchange-field effect, and hence nonzero $\Delta V_c$, may gradually appear around $d_{dead}$. The $V_c$ (or $J_c$) oscillating period of 2.6 nm and the characteristic length given by $\hbar v_F/2E_{ex} = 1$ nm roughly agree with past reports regarding simple Nb/Ni/Nb JJs[28,30,31].

Possibilities of new device applications follow from this exchange-field effect. The node in each $V_c(d_{Ni})$ in Fig. 4 represents the transition of the zero-field JJ ground state from 0 to π in phase difference[1,2,25,26]. Such transitions occur at different $d_{Ni}$ values of 1.6 nm and 2.1 nm for the P and AP states, respectively, which implies that JJs with 1.6 nm < $d_{Ni}$ < 2.1 nm are phase-switchable devices. Phase-shifting elements are novel components of superconducting digital and qubit electronics[32-35]. Among the different device types, S-F-S JJs are often considered the most promising architecture for a π-phase-shifter due to their nonvolatility and small size[35]. For example, the proposed elimination of the superconducting loops in some rapid single-flux quantum logic components by use of π-JJs[33] has been experimentally demonstrated with S-F-S devices[36]. It will be interesting to see what kind of novel future electronics will be conceived and realized with the added capability of *in situ*, nonvolatile phase-switching offered by S-PSV-S JJs. Regarding the cryogenic memory application of these devices, the near-extinction of $I_c$ at a 0-π transition is an important feature, since it facilitates reliable discrimination of the information stored in the PSV states. A large bias-current margin for discriminating between the low and high $I_c$ states in a cryogenic memory based on single JJs will be essential in overcoming device-parameter spreads in highly-integrated circuits. A useful metric for this margin is the relative change in $I_c$, $|\Delta I_c|/$(the lesser of $I_c^P$ and $I_c^{AP}$) ≈ 500 % for $d_{Ni}$ = 1.5 nm (Fig. 3c). This is well beyond a typical GMR ratio of a spin valve (< 10%)[18] and comparable to the best present-day tunneling magnetoresistance (TMR) ratios of ≈ 600 % at room temperature and ≈ 1100 % at 4.2 K[37]. Determining the fundamental limit to that margin may require investigation of higher-harmonic Josephson currents[27,30,38]. Such a large margin may allow reading memory states with a very low error rate. Also the use of Josephson energy as a memory parameter implies the inherent compatibility with SFQ (single flux quantum) electronics and may be probed by an SFQ. Because an SFQ signal is ballistic with a speed of light and only dissipates very low energy in a junction, memory elements based on switchable Josephson energy may prove to be a way to overcome the speed and power limitation of conventional charge- or resistance-based devices.

In this work, we distinguished the exchange field behavior from the remanent-field effects and showed that it is a size-independent phenomenon. However, further research on smaller nanoscale devices is needed to determine the scalability limits of these devices. More efficient magnetization switching with spin-transfer torque also may be effective in nanoscale PSVs. The results in this paper demonstrate that Josephson junctions with pseudo-spin-valve barriers have the potential to enable low-power, high-speed, high-density cryogenic memory for a high-performance superconducting computing system.

**Acknowledgements**

The authors thank M. R. Pufall and H. Rogalla for technical assistance and valuable discussions. This work was supported by NIST and by U.S. National Security Agency under agreement numbers EAO156513 and EAO176792.

a. Work of the U. S. government, not subject to U. S. copyright.

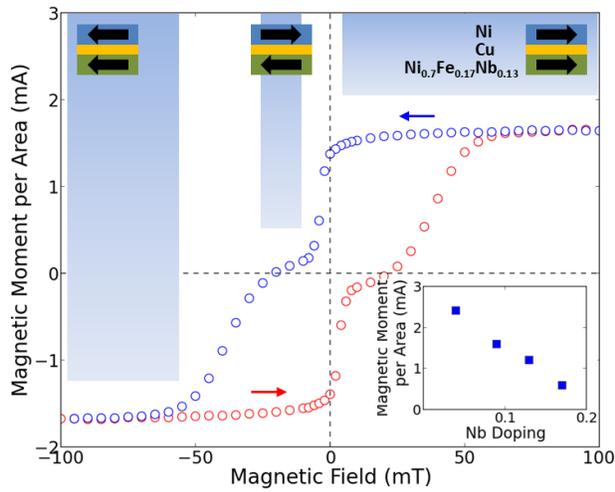

**Figure 1. Hysteretic magnetization data at 10 K from an unpatterned PSV multilayer structure: Nb(100 nm)/Cu(3 nm)/Ni$_{0.7}$Fe$_{0.17}$Nb$_{0.13}$(2.1 nm)/Cu(5 nm)/Ni(3 nm)/Cu(3 nm)/Nb(70 nm).** The field was swept from positive to negative (blue circles), then back to positive (red circles), as indicated by the colored arrows. Illustrated above the plot are the different magnetization states of the Ni and Ni$_{0.7}$Fe$_{0.17}$Nb$_{0.13}$ for the trace with the blue circles. Inset: Trend of Ni$_{0.7}$Fe$_{0.17}$Nb$_{0.13}$ saturation magnetization with Nb doping.

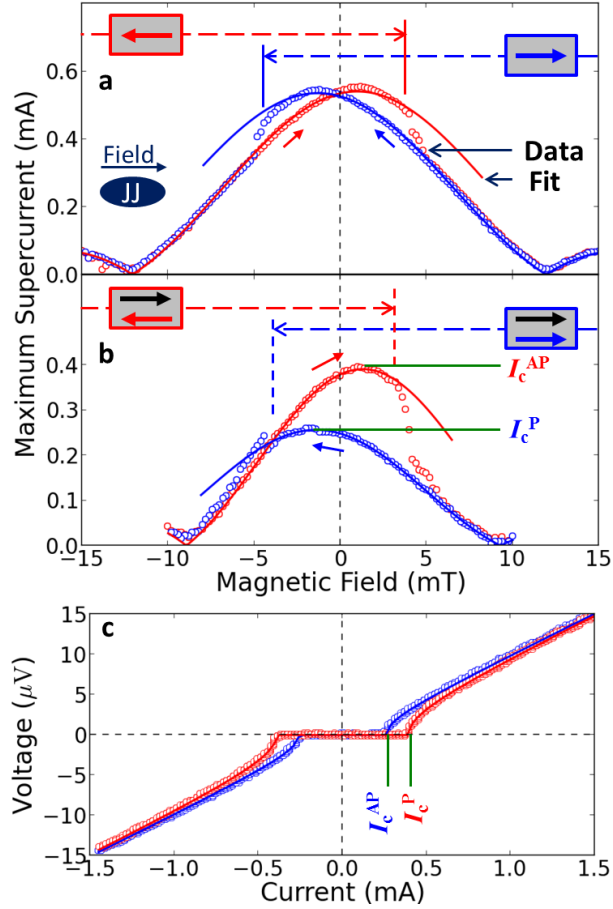

**Figure 2. Electrical measurement results of S-PSV-S JJs at 4 K. a**, Maximum supercurrent *vs.* magnetic field of a JJ with zero Ni thickness ($d_{Ni} \approx 0$) in the PSV barrier structure, $Ni_{0.7}Fe_{0.17}Nb_{0.13}$(2.1 nm)/Cu(5 nm)/Ni($d_{Ni}$). The JJ design is a 0.9 μm × 1.8 μm ellipse and the magnetic field is applied along the long axis of the ellipse (inset). **b**, Maximum supercurrent *vs.* magnetic field of the S-PSV-S device with $d_{Ni}$ = 1 nm. The JJ design is a 1.2 μm × 2.4 μm ellipse. **c**, Current *vs.* voltage of the device used in **b** measured at the magnetic field giving in the peak $I_m$ for each state. Symbols are measured data and lines are fits. The magnetization states are labeled by colored and black arrows in the illustrated boxes above the plots for the $Ni_{0.7}Fe_{0.17}Nb_{0.13}$ and the Ni layers, respectively.

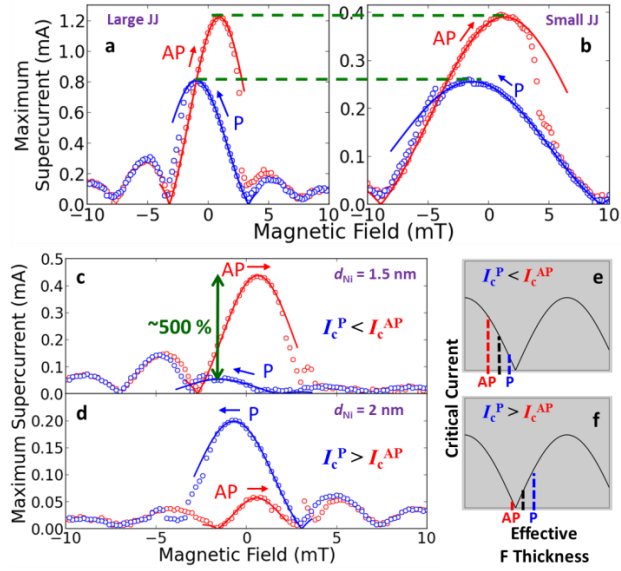

**Figure 3. Maximum supercurrent *vs*. magnetic field characteristics of S-PSV-S JJs at 4 K with different areas and Ni thicknesses. a** and **b** show the effects of different junction areas. The actual areas are estimated to be **a**, 2.6 μm² and **b**, 0.78 μm², based on the $R_nA_{eff} \approx 8.0$ mΩ μm². The JJ designs are **a**, 1.6 μm × 3.2 μm and **b**, 1.2 μm × 2.4 μm ellipses. The Ni thickness is 1 nm for both JJs. **c** and **d** show very large $\Delta I_c$ and with opposite signs for two different Ni thicknesses **c**, 1.5 nm and **d**, 2 nm. As designed, both JJs are 1.4 μm × 2.8 μm ellipses. The symbols and the curves represent data and fits, respectively. **e-f,** Illustrations of the origin of the different $\Delta I_c$ in **c** and **d**, respectively. Effective F thickness means the Ni thickness that would result in the same phase shift in the PSV. The decay in $I_c$ is ignored for simplicity. A P or AP state results in an increased or decreased phase (blue or red dashed line) relative to that given by the Ni thickness only (black dashed lines).

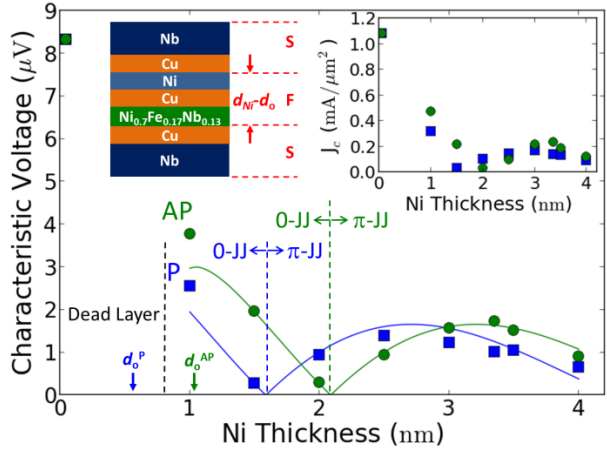

**Figure 4. Characteristic voltage $V_c$ *vs*. Ni thickness data (symbols) and fit (solid line) in the S–PSV–S devices with the PSV structure $Ni_{0.7}Fe_{0.17}Nb_{0.13}$(2.1 nm)/Cu(5 nm)/Ni($d_{Ni}$) at 4 K.** P and AP states are indicated by blue squares and green circles, respectively. Each $V_c$ datum is an average for a few JJ samples. For $d_{Ni} < 3$ nm, each $V_c$ datum is an average for either 3 or 4 devices, resulting in an error bar comparable to or smaller than the maker size (the standard error of the mean $I_c \leq 20$ % of the mean $I_c$). For $d_{Ni} \geq 3$ nm, the sample size is 1. Left inset: device multilayer structure and its equivalent S-F-S structure as an approximation. Right inset: critical current density $J_c$ (given by $V_c/R_n A_{eff}$) *vs*. Ni thickness.